%% file: hello_world.tex
\documentclass[submission,copyright,creativecommons]{eptcs}
\providecommand{\event}{TTC 2011} 

\def\titlerunning{Saying Hello World with Epsilon}
\def\authorrunning{L.M. Rose, A. Garc\'{\i}a-Dom\'{\i}nguez, J.R. Williams, D.S. Kolovos, R.F. Paige  \& F.A.C. Polack}

\usepackage{url}
\usepackage{graphicx}
\usepackage{courier}
\usepackage{listings}

\input{listings.tex}

\title{Saying Hello World with Epsilon -- A Solution to the \\ 2011 Instructive Case}

\author{Louis M. Rose$^{1}$ \and Antonio Garc\'{\i}a-Dom\'{\i}nguez$^{2}$ \and James R. Williams$^{1}$ \and Dimitrios S. Kolovos$^{1}$ \and Richard F. Paige$^{1}$ \and Fiona A.C. Polack$^{1}$ \institute{$^{1}$ Department of Computer Science, University of York, UK.} \email{[louis,jw,dkolovos,paige,fiona]@cs.york.ac.uk}  \institute{$^{2}$ University of C\'{a}diz, Department of Computer Languages and Systems\\C/Chile 1, 11002, C\'{a}diz, Spain} \email{antonio.garciadominguez@uca.es}}

\begin{document}
\maketitle

\begin{abstract}
Epsilon is an extensible platform of integrated and task-specific languages for model management. With solutions to the 2011 TTC Hello World case, this paper demonstrates some of the key features of the Epsilon Object Language (an extension and reworking of OCL), which is at the core of Epsilon. In addition, the paper introduces several of the task-specific languages provided by Epsilon including the Epsilon Generation Language (for model-to-text transformation), the Epsilon Validation Language (for model validation) and Epsilon Flock (for model migration). 
\end{abstract}

\section{Introduction}
This paper presents a solution to the 2011 TTC Hello World case that uses Epsilon, an extensible platform of integrated and task-specific languages for model transformation, validation, merging, comparison, refactoring and migration \cite{kolovos09thesis}. To provide an introduction to Epsilon for new users, each part of the solution is discussed and several of the Epsilon languages are demonstrated. Epsilon is briefly described below, and then solutions to each of the problems in the case are presented. 

\subsection{Epsilon}
Epsilon is built atop Eclipse, and interoperates seamlessly with several modelling technologies, including EMF, MDR, CZT and plain XML. Further information on Epsilon can be found on the project website\footnote{\url{http://www.eclipse.org/gmt/epsilon}} and in the Epsilon book\footnote{\url{http://www.eclipse.org/gmt/epsilon/doc/book}}. The Epsilon Object Language (EOL) \cite{kolovos06eol} is at the core of the Epsilon and provides functionality similar to that of OCL. However, EOL provides an extended feature set, which includes the ability to update models, access to multiple models, conditional and loop statements, statement sequencing, and provision of standard output and error streams.

The solutions described in this paper use EOL, as well as the Epsilon Generation Language (EGL) \cite{rose08egl} for model-to-text transformation, the Epsilon Validation Language (EVL) \cite{kolovos08evl} for model validation, and Epsilon Flock \cite{rose10flock} for a specialised form of in-place model-to-model transformation (model migration). Further languages in the Epsilon platform which were not used for the Hello World case include the Epsilon Comparison Language (ECL) for comparing and matching models, the Epsilon Merging Language (EML) for combining models, and the Epsilon Wizard Language (EWL) for refactoring models. 

\input{greetings.tex}
\input{counting.tex}
\input{reverse.tex}
\input{migration.tex}
\input{delete.tex}

\section{Opponent Statements}
Three opponents were assigned to the Epsilon solution, and their statements\footnote{\url{http://planet-research20.org/ttc2011/index.php?option=com_community&amp;view=groups&amp;task=viewdiscussions&amp;groupid=13&amp;Itemid=150&amp;view=groups&amp;task=viewdiscussions&amp;groupid=13&amp;Itemid=150} (registration required)} are now summarised. Every opponent remarked that most of the solutions to the Hello World are very concise and readable when formulated with Epsilon. However, solutions that required matching complex patterns (such as finding cycles of three nodes in a graph, Section~\ref{sec:counting}) were less concise and readable due to the use of imperative constructs for specifying patterns. We are investigating the possibility of adding pattern matching constructs to EOL, via a more declarative style of syntax. 

Two of the statements remarked that using a family of task-specific languages enhanced readability and conciseness of solutions, and the understandability of Epsilon as a whole. One of the statements suggested that learning the similarities and differences between the family of languages might be a challenge for new users of Epsilon. To smooth and reduce the learning curve for users, Epsilon provides online documentation\footnote{\url{http://www.eclipse.org/gmt/epsilon/doc/}} including examples of Epsilon programs, tutorial articles, and even a free book\footnote{\url{http://www.eclipse.org/gmt/epsilon/doc/book}}. Finally, as one of the statements suggests, we are currently working on extending the content assistance provided by the Epsilon development tools to support, for example, auto-completion for metamodel types.

\bibliography{hello_world}
\bibliographystyle{eptcs}

\newpage

\appendix
\input{optional_tasks.tex}

\end{document}

%% file: listings.tex
\lstdefinelanguage{EOL}{
morekeywords={import,for,while,in,and,or,self,operation,return,def,var,throw,if,new,else,transaction,abort,break,continue,assert,assertError,not,delete},
sensitive=true,
morecomment=[l]{--},
morestring=[b]',
showstringspaces=false
}

\lstdefinelanguage{EGL}{
morekeywords={import,for,while,in,and,or,self,operation,return,def,var,throw,if,new,else,transaction,abort,break,continue,assert,assertError,not,out},
sensitive=true,
morecomment=[l]{--},
morestring=[b]',
showstringspaces=false
}

\lstdefinelanguage{EVL}
{morekeywords={for, import, context, typeOf, kindOf, high, medium, low, constraint, critique, check, do, message, title, fix, not, guard, in, and, or, operation, return, var, def, if, new, else, delete},
sensitive=true,
morecomment=[l]{--},
morestring=[b]',
showstringspaces=false,
}

\lstdefinelanguage{Flock}
{morekeywords={for, import, migrate, to, when, original, migrated, typeOf, kindOf, do, not, in, and, or, operation, return, var, if, new, else, delete},
sensitive=true,
morecomment=[l]{--},
morestring=[b]',
showstringspaces=false,
}

%% file: greetings.tex

\section{Task 1: Greeting with EOL and EGL}
EOL has been used to instantiate the Hello World metamodels \cite[Section 2.1]{case}. In EOL, model elements are created using the \texttt{new} keyword and a model element's attributes are accessed with dot notation. The EOL program in Listing~\ref{lst:simple_hello_world} instantiates the \texttt{Greeting} metamodel type (line 1) and sets the \texttt{text} feature of the \texttt{Greeting} (line 2). References are accessed in the same manner as attributes: using dot notation (see lines 4 and 7 of Listing~\ref{lst:extended_hello_world}), for example.

EOL is dynamically typed. Consequently, EOL programs do not specify to which metamodel particular types belong (e.g. \texttt{Greeting}). Instead, the user can specify the models (and hence metamodels) on which the EOL program operates just before executing the program. When using the Eclipse-based development tools, for example, the user specifies the name, type, location and metamodel of each model when creating an Eclipse launch configuration.

\begin{lstlisting}[caption=Instantiating the simple Hello World metamodel with EOL., label=lst:simple_hello_world, language=EOL]
var g = new Greeting;
g.text = "Hello World";
\end{lstlisting}

\begin{lstlisting}[caption=Instantiating the extended Hello World metamodel EOL., label=lst:extended_hello_world, language=EOL]
var g = new Greeting;

g.greetingMessage = new GreetingMessage;
g.greetingMessage.text = "Hello";

g.person = new Person;
g.person.name = "TTC Participants";
\end{lstlisting}

For model-to-text transformation, Epsilon provides a dedicated task-specific language, the Epsilon Generation Language (EGL) \cite{rose08egl}. EGL is a template-based language that provides dynamic and static sections. The former contain EOL code which is used to extract data from input models; while the latter contain plain text which remains constant for varying input models. Dynamic sections can emit text using the \texttt{out.print} operation. Alternatively, the \texttt{[\%=text\%]} syntactic shortcut can be used (instead of \texttt{[\% out.print(text); \%]}).

Listing~\ref{lst:m2t_hello_world} uses three dynamic sections (contained in \texttt{[\% \%]} tags) and two static sections (containing a space and an exclamation mark, respectively, on line 2). The dynamic section on line 1 uses the \texttt{all}  (available on every metamodel type and used to retrieve all of the instances of the that metamodel type) and \texttt{first} (used to retrieve the first element of a collection) properties to find the sole instance of \texttt{Greeting} in the input model. The two dynamic sections on line 2 emit the value of the \texttt{greetingMessage.text} and \texttt{person.name} features of the \texttt{Greeting}, respectively. Executing the template in Listing~\ref{lst:m2t_hello_world} emits a string of the form \texttt{<text> <name>!} (such as \texttt{Hello Franz!}).

\begin{lstlisting}[caption=A model-to-text transformation with EGL., label=lst:m2t_hello_world, language=EGL]
[% var g = Greeting.all.first; %]
[%=g.greetingMessage.text%] [%=g.person.name%]!
\end{lstlisting}

%% file: counting.tex

\section{Task 2: Counting with EOL}
\label{sec:counting}
The \texttt{all} property (discussed above) returns a collection containing all of the instances of the specified metamodel type. In EOL, the number of elements in a collection is determined using the \texttt{size} property. Hence, line 1 of Listing~\ref{lst:simple_counting} prints the number of \texttt{Node}s in the input model. Like OCL, EOL provides a number of higher-order operations for collection types. The \texttt{select} operation, for example, returns a filtered copy a collection containing only those elements that satisfy the specified predicate. Therefore, line 2 of Listing~\ref{lst:simple_counting} prints the number of \texttt{Edge}s whose \texttt{src} and \texttt{trg} features are equal (i.e. looping edges). Line 3 performs a similar query for counting isolated nodes using a \emph{user-defined operation}.

User-defined operations allow existing (primitive or metamodel) types to be enriched with additional functionality. For example, lines 5-7 of Listing~\ref{lst:simple_counting} define an operation, \texttt{isIsolated}, for the \texttt{Node} type. The \texttt{isIsolated} operation returns a \texttt{Boolean} value. The body of the \texttt{isIsolated} operation uses another of EOL's higher-order operations for collections, \texttt{exists}, which implements existential quantification. 

\begin{lstlisting}[caption={Counting nodes, looping edges and isolated nodes with EOL.}, label=lst:simple_counting, language=EOL]
Node.all.size.println();
Edge.all.select(e|e.src == e.trg).size.println();
Node.all.select(n|n.isIsolated()).size.println();

operation Node isIsolated() : Boolean {
	return not Edge.all.exists(e|e.src == self or e.trg == self); 
}
\end{lstlisting}

Finding patterns of more than one or two model elements in EOL is complicated to specify in terms of the higher-order operations on collections, and hence imperative programming constructs are typically used instead. For example, finding cycles of three \texttt{Node}s involves performing a depth-first search over the successors of each \texttt{Node} in the model (Listing~\ref{lst:counting_circles}). The \texttt{successors()} operation is not shown, for brevity, but traverses the outgoing \texttt{Edge}s of a \texttt{Node} to identify successor \texttt{Node}s.

\begin{lstlisting}[caption=Counting circles of three nodes in EOL., label=lst:counting_circles, language=EOL]
var results : Sequence;

for (node1 in Node.all) {
  for (node2 in node1.successors()) {
    if (node2 == node1) continue;
    for (node3 in node2.successors()) {
      if (node3 == node2 or node3 == node1) continue;
      if (node3.successors().contains(node1)) {
        results.add(Sequence { node1, node2, node3 });
      }
    }
  }
}
results.println();
\end{lstlisting}

Section~\ref{subsec:dangling_edges} describes the optional task of checking for dangling edges.

%% file: reverse.tex

\section{Task 3: Reversing with EOL and Epsilon Flock}
\label{sec:reverse}
Epsilon programs are granted read-only, write-only or read-write access to a particular model by the user. As such, EOL can be used for querying models (read-only), constructing models (write-only) or modifying models (read-write). For reversing all of the \texttt{Edge}s in a model, EOL has been used to specify an in-place update transformation (i.e. read-write access) on the input model (Listing~\ref{lst:reverse}). The \texttt{for} construct is used to iterate over the \texttt{Nodes} in the input model. EOL does not support parallel assignment, so a temporary (\texttt{temp}) is used. 

\begin{lstlisting}[float=tbp, caption=Reversing edges with EOL., label=lst:reverse, language=EOL]
for (edge in Edge.all) {
	var temp = edge.src;
	edge.src = edge.trg;
	edge.trg = temp;
}
\end{lstlisting}

Alternative, an Epsilon Flock \cite{rose10flock} migration strategy can be used to reverse edges. Compared to the EOL solution (Listing~\ref{lst:reverse}), the iteration is performed declaratively (\texttt{migrate Edge} on line 1 of Listing~\ref{lst:reverse_flock}) rather than imperatively (with a \texttt{for} loop), and no temporary variable is required because Flock provides the \texttt{original} and migrated \texttt{model} elements, which are bound to distinct model elements. Epsilon Flock is discussed further in Section~\ref{sec:migrating}.

\begin{lstlisting}[caption=Reversing edges with Flock., label=lst:reverse_flock, language=Flock]
migrate Edge {
	migrated.src = original.trg.equivalent();
	migrated.trg = original.src.equivalent();
}
\end{lstlisting}

%% file: migration.tex

\section{Task 4: Migrating with Epsilon Flock}
\label{sec:migrating}
Epsilon provides a dedicated task-specific language for performing model migration, Epsilon Flock \cite{rose10flock}. Metamodel evolution typically involves changes to a small proportion of a metamodel \cite{sprinkle03thesis}, and Flock exploits this by automatically copying model elements that have not been affected by metamodel evolution. Migration rules are specified only for those model elements that have been affected by metamodel evolution. For example, the case description describes a metamodel evolution in which the \texttt{Graph} type merges its \texttt{nodes} and \texttt{edges} reference to form a new \texttt{gcs} reference, and the \texttt{name} property of the \texttt{Node} class is renamed to \texttt{text}. 

Flock migration rules are specified on a particular source metamodel type. Each rule is executed once for each instance of that type in the source model. In the body of a rule, the \texttt{original} and \texttt{migrated} variables are bound to an element of the source model and its equivalent element in the target model, respectively. Listing~\ref{lst:extract_superclass} demonstrates two migration rules: the first (lines 1-4) copies the contents of the \texttt{nodes} and \texttt{edges} references into the \texttt{gcs} reference for instances of \texttt{Graph}; and the second (lines 6-8) copies the value of the \texttt{name} feature into the \texttt{text} for instances of \texttt{Node}.

\begin{lstlisting}[caption=Migrating to the evolved graph metamodel with Flock., label=lst:extract_superclass, language=Flock]
migrate Graph {
	migrated.gcs.addAll(original.nodes.equivalent());
	migrated.gcs.addAll(original.edges.equivalent());
}

migrate Node {
	migrated.text = original.name;
}
\end{lstlisting}

Section~\ref{subsec:second_migration} describes the optional task of migrating to the even more evolved graph metamodel.

%% file: delete.tex

\section{Task 5: Deleting with EOL}
As discussed in Section~\ref{sec:reverse}, EOL programs can be used to perform in-place update transformations. Deleting a model element is possible with the \texttt{delete} keyword. Deleting the node with name \texttt{n1} (Listing~\ref{lst:delete}) has been achieved using the \texttt{selectOne} higher-order operation (a shorthand for \texttt{select} followed by \texttt{first}) to locate the relevant node, and using the \texttt{delete} keyword.

\begin{lstlisting}[caption=Deleting a node with EOL., label=lst:delete, language=EOL]
delete Node.all.selectOne(n|n.name == "n1");
\end{lstlisting}

Section~\ref{subsec:delete_node_and_edges} describes the optional task of deleting a node and its edges.

%% file: optional_tasks.tex

\section{Optional Tasks}
Solutions to the optional tasks of the case are now described.

\subsection{Task 2.5: Checking for dangling edges with EVL}
\label{subsec:dangling_edges}
Counting the dangling edges in a model can be formulated in the same manner as counting isolated nodes (Section~\ref{sec:counting}). However, the Epsilon Validation Language (EVL) \cite{kolovos08evl} can be used to check for -- and reconcile -- dangling edges by specifying a validation constraint and a corresponding \texttt{fix} (an in-place transformations that reconcile validation problems), as shown in Listing~\ref{lst:dangling_edges_evl}. An EVL \texttt{constraint} (line 2) is specified in the \texttt{context} (line 1) of a particular metamodel type (\texttt{Edge} in Listing~\ref{lst:dangling_edges_evl}). When the \texttt{check} (line 3) part of a constraint is not satisfied (returns \texttt{false}), the user is presented with the \texttt{message} part (line 4) of the constraint and can optionally invoke one of the \texttt{fix} parts (lines 5-11). The constraint in Listing~\ref{lst:dangling_edges_evl} provides one \texttt{fix}, which deletes the dangling edge from the model. EVL executes the \texttt{DanglingEdges} constraint once for every instance of \texttt{Edge} in the input model.

\begin{lstlisting}[caption=Checking for dangling edges with EVL., label=lst:dangling_edges_evl, language=EVL]
context Edge {
	constraint DanglingEdges {
		check: not self.isDangling()
		message: "The edge " + self + " is dangling."
		fix {
			title: "Remove this edge"
			
			do {
				delete self;
			}
		}
	}
}

operation Edge isDangling() : Boolean {
	return self.src.isUndefined() or self.trg.isUndefined(); 
}
\end{lstlisting}

\subsection{Task 4.2: Migrating to the even more evolved graph metamodel with Epsilon Flock}
\label{subsec:second_migration}
The case describes a second metamodel evolution in which edges are specified with reference values rather than model elements. The evolved metamodel no longer contains an \texttt{Edge} class, and instead the \texttt{Node} class references itself via the \texttt{linksTo} reference. Listing~\ref{lst:second_migration} demonstrates the way in which this migration can be specified for Epsilon Flock.

Flock provides an Eclipse extension point for distributing migration strategies to metamodel users. The current version of Flock (0.9) does not provide built-in support for chaining multiple migration strategies together, but this could be achieved by using Java to iterate over each migration strategy file, invoking Flock for each strategy.

\begin{lstlisting}[caption=Migrating to the even more evolved graph metamodel with Flock., label=lst:second_migration, language=Flock]
migrate Graph {
	migrated.nodes = original.gcs.equivalent();
}

migrate Node {
	migrated.linksTo = original.successors().equivalent();
}

operation Original!Node successors() : Collection(Node) {
	return self.outgoing().collect(e|e.trg);
} 

operation Original!Node outgoing() : Collection(Edge) {
	return Edge.all.select(e|e.src == self);
}
\end{lstlisting}

\subsection{Task 5.2: Removing a node and its incident edges with EOL}
\label{subsec:delete_node_and_edges}
The \texttt{delete} keyword removes from the model a model element and all model elements contained in the deleted model element. To delete a \texttt{Node} and its incident \texttt{Edge}s, three delete keywords have been used (Listing~\ref{lst:delete_node_and_edges}) because, in the graph metamodel provided by the case, a \texttt{Node} does not contain its \texttt{Edge}s.

\begin{lstlisting}[caption=Deleting a node and its incident edges with EOL., label=lst:delete_node_and_edges, language=EOL]
var n1 : Node = Node.all.selectOne(n|n.name == "n1");

delete n1.incoming();
delete n1.outgoing();
delete n1;

operation Node incoming() : Collection(Edge) {
	return Edge.all.select(e|e.trg == self);
}

operation Node outgoing() : Collection(Edge) {
	return Edge.all.select(e|e.src == self);
}
\end{lstlisting}

Alternatively, a Flock migration strategy can be used to specify the nodes and edges that should not be copied to the migrated model (Listing~\ref{lst:delete_node_and_edges_flock}). Flock provides the \texttt{delete} construct for specifying model elements that should not be copied. Deletions are guarded using the \texttt{when} keyword. 

\begin{lstlisting}[caption=Deleting a node and its incident edges with EOL., label=lst:delete_node_and_edges_flock, language=Flock]
delete Node when: original.name == "n1"

delete Edge when: original.src.name == "n1" or
                        original.trg.name == "n1"
\end{lstlisting}

\subsection{Task 6: Inserting transitive edges with EOL}
Inserting transitive edges with EOL is a two-step process (Listing~\ref{lst:transitive}). First, the graph is inspected to calculate the set of new edges to be created (using the \texttt{calculateTransitiveEdges} operation on lines 7-21). The \texttt{src} and \texttt{trg} \texttt{Nodes} are stored in the set for each transitive edge to be created. Secondly, an instance of \texttt{Edge} is created for each member of the set (using the \texttt{calculateTransitiveEdges} operation on lines 23-31). The two steps are separate to ensure that transitive edges are created from only those edges that existed at the start of the transformation, and not for edges added partway through the transformation.

Notice that the \texttt{successors} and \texttt{outgoing} operations on (lines 33-36 and 38-41 respectively) might be called more than once for each Node in the model. Consequently the \texttt{@cached} annotation is used to indicated that EOL should cache the results of executing these operations. Caching provides a mechanism for increasing execution time when the result of an operation remains constant throughout the execution of an Epsilon program.

\begin{lstlisting}[caption=Inserting transitive edges with EOL., label=lst:transitive, language=EOL]
var g         = Graph.all.first;
var edgeSpecs = g.calculateTransitiveEdges();

g.addEdgesFromSpecification(edgeSpecs);

operation Graph calculateTransitiveEdges() : Set {
	var edgeSpecs = new Set;

	for (node in Node.all) {
		for (successor in node.successors()) {
			for (grandsuccessor in successor.successors()) {
				if (not Edge.all.exists(e|e.src == node and e.trg == grandsuccessor)) {
					edgeSpecs.add(Sequence { node, grandsuccessor });
				}
			}
		}
	}
	
	return edgeSpecs;
}

operation Graph addEdgesFromSpecification(edgeSpecs : Set) {
	for (edgeSpec in edgeSpecs) {
		var edge : Edge = new Edge;
		edge.src = edgeSpec.first;
		edge.trg = edgeSpec.second;
		self.edges.add(edge);
		(edge.src.name + "->" + edge.trg.name).println();
	}
}

@cached
operation Node successors() : Collection(Node) {
	return self.outgoing().collect(e|e.trg);
}

@cached
operation Node outgoing() : Collection(Edge) {
	return Edge.all.select(e|e.src == self);
}
\end{lstlisting}